\pdfoutput=1
\documentclass[12pt,preprint]{aastex}

\def\kms{kms$^{-1}$\space}
\def\micron{$\mu$m\space}
\def\arcsec{$^{\prime\prime}$\space}

\shorttitle{Highly Collimated Jets in HH46/47}
\shortauthors{Velusamy et al.}

\begin{document}

\title{Highly Collimated Jets and Wide-Angle Outflows in HH46/47:\\
New Evidence from Spitzer IR Images }

\author{T. Velusamy\altaffilmark{1},William D. Langer\altaffilmark{1}, Kenneth. A. Marsh\altaffilmark{1,2}}

\altaffiltext{1}{Jet Propulsion Laboratory, Pasadena, CA 91109;
velusamy@jpl.nasa.gov} \altaffiltext{2}{IPAC, Caltech,  Pasadena, CA
91125}


\begin{abstract}
We present new details of the structure and morphology of the jets
and outflows in HH46/47 as seen in Spitzer infrared images from IRAC
and MIPS, reprocessed using the ``HiRes'' deconvolution technique.
HiRes improves the visualization of spatial morphology by enhancing
resolution (to sub-arcsec  levels in  IRAC bands) and removing the
contaminating side lobes from bright sources.  In addition to
sharper views of previously reported
 bow shocks, we have detected: (i) the sharply-delineated cavity walls of the
wide-angle biconical outflow, seen in  scattered light on both sides
of the protostar, (ii) several very narrow jet features
 at distances $\sim$400 AU to $\sim$0.1 pc from the star, and,
 (iii) compact  emissions at MIPS 24
\micron  coincident with the jet heads, tracing the hottest
atomic/ionic gas in  the bow shocks. Together the IRAC and MIPS
images provide a more complete picture of the bow shocks, tracing
both the molecular and atomic/ionic gases, respectively. The narrow
width and alignment of all jet-related features indicate a high
degree of jet collimation and low divergence (width of $\sim$400 AU
increasing by only a factor of 2.3 over 0.2pc). The morphology of
this jet, bow shocks, wide angle outflows, and the fact that jet is
non-precessing and episodic, constrain the mechanisms for producing
the jet's entrained molecular gas, and  origins of the fast jet, and
slower wide-angle outflow.
\end{abstract}


\section{Introduction}

Protostellar jets and winds originate close to the surface of the
forming star and interact with the dense envelope surrounding the
protostar-disk system (K\"{o}nigl \& Pudritz 2000; Shu et al. 2000).
Both wide angle outflows and collimated jets from young protostars
play an important role in how stars form, as they provide a means
for protostellar disks to shed material and angular momentum, thus
regulating stellar mass via accretion of disk material onto the
star.  HH46/47 is a remarkable example of a system with jets and bow
shock cavities, and, as reported here, wide angle outflow cavities,
offering a rich observational insight into the various mechanisms at
play (Reipurth \& Heathcote 1991; Chernin \& Masson 1991;
Eisl\"{o}ffel et al. 1994, Combet et al. 2006). The jets in HH46/47
are bright and show a classic structure of a collimated flow with
several large bow shocks (HH47A, HH47C, HH47D). Recently, the
HH46/47 system was observed by Spitzer with the IRAC, IRS, and MIPS
instruments as part of the Early Release Observations, and the
results were presented by Noriega-Crespo et al. (2004) and Raga et
al. (2004).  In the IRAC bands, they clearly detect the bow shock
and its cavities; these IR emissions arise from the the H$_2$
rotational lines and possibly some contribution from PAHs. However,
in these images many of the features in both the IRAC and MIPS data
are overwhelmed by the diffraction lobes from the very bright
central source.  Here we present new results for HH46/47 based on
reprocessing the IRAC and MIPS Spitzer archive data using a
deconvolution algorithm. The ``HiRes'' deconvolution  developed for
Spitzer images by Backus et al. (2005) is based on the
Richardson-Lucy algorithm (Richardson 1972; Lucy 1974), and the
Maximum Correlation Method (Aumann et al. 1990) used for IRAS data.
HiRes deconvolution improves the visualization of spatial morphology
by enhancing resolution (to sub-arcsec levels in the IRAC bands) and
removing the contaminating sidelobes from bright sources (Velusamy
et al. 2007a \& 2007b).

\section{Results}

In Figs. 1-3, we present the HiRes deconvolved images of HH46/47 in
all IRAC bands and in MIPS 24 $\mu$m using the data in Spitzer
archives. For comparison the diffraction-limited ``mosaic'' images
are also shown in Fig. 1.  The effects of the resolution enhancement
(e.g. from 2\farcs4 to $\sim$ 0\farcs8 at 8$\mu$m) and the removal
of the diffraction lobes (with residues near the Airy rings at a
level $<$ 0.05\% of the peak intensity) are clearly evident in these
images. A wide-angle outflow cavity is now clearly detected in
scattered light in the 3.6 and 4.5 \micron images, while the bow
shocks and their limb brightened cavities appear prominently in all
IRAC bands. At 24 $\mu$m, we detect only the central protostar and
two compact sources near the tip of the bow shocks ( Figs. 1 \& 3).
In Fig. 2 we show a blow up of the emissions in IRAC channels 1-2
near the protostar. A composite view of the overall morphology and
identification of the individual components of the HH46/47
outflow/jet system is shown in Fig. 3.  SEDs of selected features
are shown in Fig. 4.

{\bf 2.1. Collimated jets:} The most significant result of our HiRes
processing  is the jet morphology.  In the SW, we observe
well-collimated jet features on all distance scales from the
protostar, up to its termination at the head of the bow shock HH47C
(24\micron hotspot).  In the NE the most remarkable jet feature is
the 24\micron hotspot at the jet impact site in the bow shock HH47A.
Here we do not observe any other jet emission features toward the
bowshock, unlike the case of the SW jet.  The emission features
along the SW jet characterize its collimation and quantify its
divergence: (i) closest to the protostar it appears as a protrusion
representing the entrained molecular jet closest to its base (Fig.
2).  (ii) a remarkably very narrow $\sim$1\arcsec -- 1\farcs5 wide,
and $\sim$ 10\arcsec long ``compact jet''  (Figs. 2 \& 3), is seen
at 15\arcsec from the protostar. Its long axis is perfectly aligned
with the star and 24 \micron hotspot (Fig. 3). This feature is also
visible in the H$_2$ images of Eisl\"{o}ffel et al (1994). Our HiRes
IRAC images fully resolve the jet along, and perpendicular to, its
velocity axis (Fig. 2c). We estimate a  radius of $\sim$300 AU for
the entrained molecular jet at  9000AU from the star (at distance
450 pc). (iii) a feature $\sim$10\arcsec (4,500 AU) from the star
which is coincident with an optical knot in the HST NIC2 image
(Reipurth et al. 2000) and a compact feature in the H$_2$ image
(Eisl\"{o}ffel et al. 1994). (vi) a feature at $\sim$50\arcsec (0.1
pc) from the star is aligned well with the SW jet although we cannot
exclude that it is a part of the bow shock seen in projection.

  The most prominent jet
features in the HiRes 24 \micron image are the ``hotspots'' in the
SW \& NE bow shocks. These emissions were noted in the mosaic images
by Noriega-Crespo et al (2004). Here we discuss their morphology
using the HiRes deconvolution and their relationship to the jets and
bow shocks. In the IRAC bands the emissions from the bow shocks
delineate  long arcs that extend to large distances backwards from
the jet heads to the protostar. The emissions  in the IRAC bands are
considered to be from pure rotational H$_2$ lines excited by
C-shocks (Neufeld et al. 2006). In contrast, the MIPS 24\micron
emissions appear to be coincident with the head of the jets,
representing emission generated by shocks in a Working Surface (WS).
They are compact and somewhat more extended perpendicular to the
jet$\sim$2\farcs2$\times$2\farcs8 and 3\farcs0$\times$5\farcs9 axis,
sizes for the SW (HH47C) and NE (HH47A) hotspots, respectively. The
morphology of the 24\micron emission (possibly from the J-shock in
the WS) is distinctly different from that associated with the
molecular shocks (C-shocks) extending further back. The SEDs of
these hotspots show clear  24 \micron excess in contrast to that
along the bow shock (Fig. 4).  The optical (Hartigan et al. 1990)
and IR (Noriega-Crespo et al. 2004) spectra in HH47A  show only
strong atomic/ionic lines and no continuum.  We can therefore rule
out the MIPS 24\micron emission as due to warm dust and instead it
originates in the hottest gas in the bow-shock. In the IRS spectrum
of HH47A (Fig. 6 in Noriego-Crespo et al. 2004) bright [FeII] line
emissions at 24.51 \micron and 25.98 \micron lie within the MIPS 24
\micron passband. Indeed, the faint emission ($\sim$ 10 mJy) in the
MIPS 24 \micron image is consistent with these line intensities
considering the respective widths of the lines and the MIPS 24
\micron pass band. Thus both the NE and SW 24 \micron hotspots can
be regarded as tracing the hottest atomic/ionic gas in the bow
shocks and therefore they identify the current impact location of
the jet with the ambient envelope.

We use the multiple emission features along the SW jet to estimate
the divergence of the jet.  The compact jet at 20\arcsec (9,000 AU)
from the star has a width of 1\farcs1  perpendicular to the jet and
the hotspot at a distance of 116\arcsec (0.25 pc) has a width of
2\farcs8. Taking into consideration the HiRes beams  we estimate the
divergence of the jet entrained molecular material is a factor of
$\sim$2.3 over a distance 96\arcsec ($\sim$ 0.21pc). This divergence
of the jet (an increase in radius, $\sim$290 AU at 24 \micron
hotspot), indicates an expansion at 1.8 \kms perpendicular to the
jet, assuming a typical jet velocity of $\sim$250 \kms
(Eisl\"{o}ffel,  \& Mundt,  1994).  This expansion velocity is
consistent with free expansion of the molecular gas transverse to
the jet axis at the co-moving sound speed ($\sim$ 2\kms for H$_2$ at
T$_{gas}$ $\sim 10^{3}$ K). Thus, assuming no external confinement,
the jet has remained highly collimated over $\sim$ 0.2 pc (up to its
impact with the surrounding envelope). The  spatially discrete
emissions along the jet axis suggests the jet activity is  episodic
and is consistent with the larger parsec scale outflows observed in
HH46/47 (Stanke et al. 1999). The alignment of  the knots, and the
jet head and the linear morphology of the compact jet rules out a
precessing jet.


{\bf 2.2. Wide-angle outflows:} The wide angle outflows are clearly
detected in the 3.6 and 4.5 \micron HiRes images (Figs. 1 - 3) and
this is the first evidence of a bipolar (NE \& SW) wide-angle
outflow cavity in HH46/47 as observed in the scattered light from
the protostar. The optically observed parabolic sheath of reflection
nebulosity towards the NE (Reipurth et al. 2000) traces parts of
this outflow, which is also observed in all IRAC bands.  However,
the scattered light at 3.6 and 4.5 \micron traces a much wider
outflow (Figs. 2 \& 3). The NE outflow lobe is clearly a mix of an
inner bright parabolic cavity traced in both reflected light and
entrained molecular gas, and a broader wide-angle  component
detected only in scattered light at 3.6 $\mu$m.  The wide-angle
outflow cavity towards the SW is not detected in the optical,
possibly due to its faintness and the viewing geometry  and it may
be hidden behind ambient cloud material (Stanke et al. 1999).  With
this new detection of the wide-angle outflow to the SW (apparent
opening angle of $\sim$110$^{\circ}$), HH46/47 emerges as a
classical example of an wide-angle outflow - narrow jet system,
placing it among those with very wide outflows (Velusamy \& Langer
1998; Arce \& Sargent 2006). It would be interesting if such wide
opening angles in HH46/47 is the result of outflow evolution with
age as proposed by Velusamy \& Langer (1998).   Since outflows can
disperse the envelope and modify the infall geometry, this provides
a natural mechanism to stop the infall, ending the accreting phase.
\section{Discussion}
Our HiRes deconvolved Spitzer images of HH46/47 show the concurrent
presence of both highly collimated fast bipolar jets and poorly
collimated slower wide-angle outflow.  The cool molecular gas in the
wide-angle outflow (because it does not emit in other IR bands,
unlike the jet or the bow shock) is observed along with the presence
of warmer entrained molecular gas along the jet close to the
protostar (within 1\arcsec--2$^{\prime\prime}$). The
magnetocentrifugal origin of jets and their launch from the
magnetized accretion disk of the protostar (Ouyed \& Pudritz 1997)
are generally accepted, although the detailed mechanism is under
debate.  In contrast, the wide angle outflow may be jet driven (Raga
\& Cabrit 1993, Ostriker et al. 2001) or wind driven (Shu et al.
2000).  In the jet driven model the bow shock produces a thin shell
that stretches out from the jet head back to the star (Fig. 3).  The
models of Raga et al. (2004) for HH47C have a leading bow shock with
wings extending all the way back to the protostar producing a dense
shell of material.  However, this bow shock bubble is much narrower
than the wide-angle outflow traced by scattered light (Fig. 3).
Clearly, the wide angle outflow is not part of this bow shock and we
can rule out a jet-driven origin.  The wide angle outflow alone can
be explained by wind-driven models. Another possible scenario for
the jet and outflow structure in HH46/47 is an MHD self-similar
model (Lery 2003, Combet et al. 2006) where radiation and
magnetocentrifugal acceleration and collimation produce heated
pressure-driven outflows.  Here, in addition to a central
accretion-ejection engine driving the atomic jet, the wide-angle
molecular outflow is powered by the infalling matter that follows a
circulation pattern around the central object without necessarily
being entrained by the jet (Figs. 4 \& 7 in Combet et al. 2006). The
molecular outflows appear as a wider hollow conical structure. The
central jet is atomic and occupies the axial region. In this
scenario, there is no strong relation between the fast jet and the
slower molecular wide angle outflow near the protostar.

In HH46/47 the SW jet is propagating inside a bubble (created by the
bow shock) and there is little chance for molecular gas entrainment
along its path.   Therefore, the detection of the jet features in
the molecular gas (IRAC bands) all along the jet axis towards the
bow shock, indicate that substantial molecular gas entrainment in
the jet must have occurred right at the base of the jet.  We
estimate a size $\sim$ 300 AU for the entrained molecular jet at its
base near the star, obtained by extrapolating back to the star the
divergence of the molecular jet  between the compact jet and the jet
head at HH47C, (section 2.1). In other words, the atomic jet which
originates within a few AU of the star must have entrained the
molecular gas within the star-disk-infall interface region which is
at least 300 AU across. Thus, our results constrain the density
structure above the base of the atomic jet in such a way that
sufficient molecular material is still available to the jet to
entrain molecular gas across a few hundred AU before it enters the
cavity created by the wide-angle outflow and bowshocks. The
molecular gas entrainment along the atomic jet occurs at least up to
a few hundred AU from the star into the star-disk-infall interface.
Such gas entrainment by the jet is possible, for example, in the
models by Combet et al (2006)  where the infall-outflow circulation
provides a molecular gas buffer over a few hundred AU across and
above the star.   This material near the vertex of the outflow cones
is sufficient for the atomic jet emerging from below to entrain
molecular gas as observed in HH46/47.

\acknowledgments

We thank the referee for helpful suggestions regarding the
interpretation of the 24 \micron emission. The research described in
this paper was carried out at the JPL, Caltech, under a contract
with NASA.  We thank Timothy Thompson for his help with the data
analysis.


\clearpage

\begin{figure}
\epsscale{1.0} \plotone{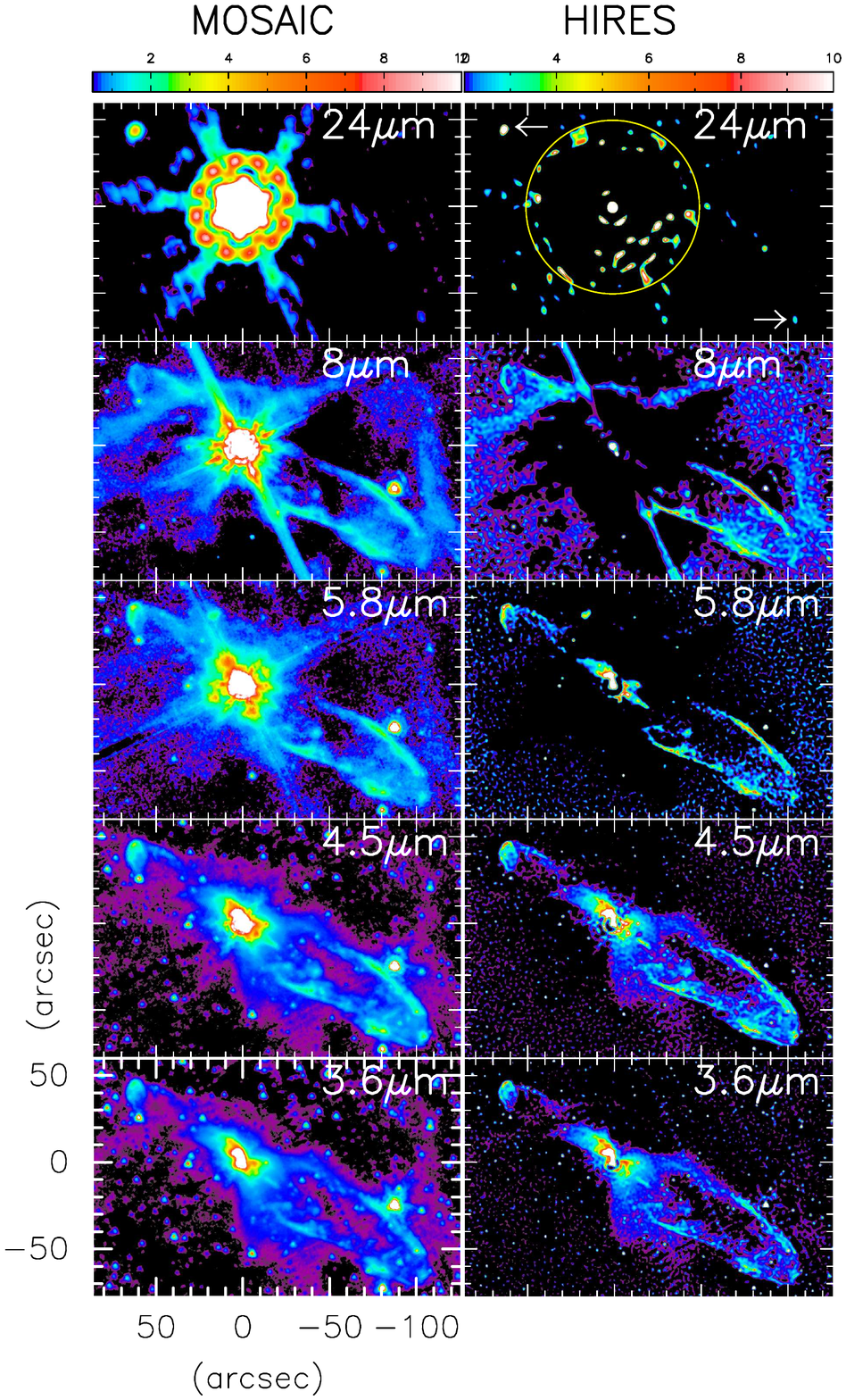} \caption{ The Mosaic (left) and
HiRes deconvolved (right) Spitzer IRAC and MIPS images of HH46/47.
The intensities are in units of MJysr$^{-1}$. Identical square root
color stretch is used for all bands such that the low surface
brightness is highlighted. The brightest emission in this display
($\sim$ 10 MJysr$^{-1}$) is $<$ 1\% of the peak in the HiRes images.
FWHM of a point source in the HiRes IRAC channels 1-4 is 0\farcs55 -
0\farcs75 and 1\farcs6 for MIPS 24 $\mu$m. The (0, 0) position is
RA(2000):08:25:43.68; Dec(2000): -51:00:34.92.  In the 8\micron
image, the NE-SW streak is an artifact. In the 24\micron HiRes
image, the region inside the circle is confused by diffraction
residue below 0.1 \% of the peak intensity. The white arrows mark
the hotspots. \label{fig1}}
\end{figure}
\clearpage

\begin{figure}
\epsscale{0.75} \plotone{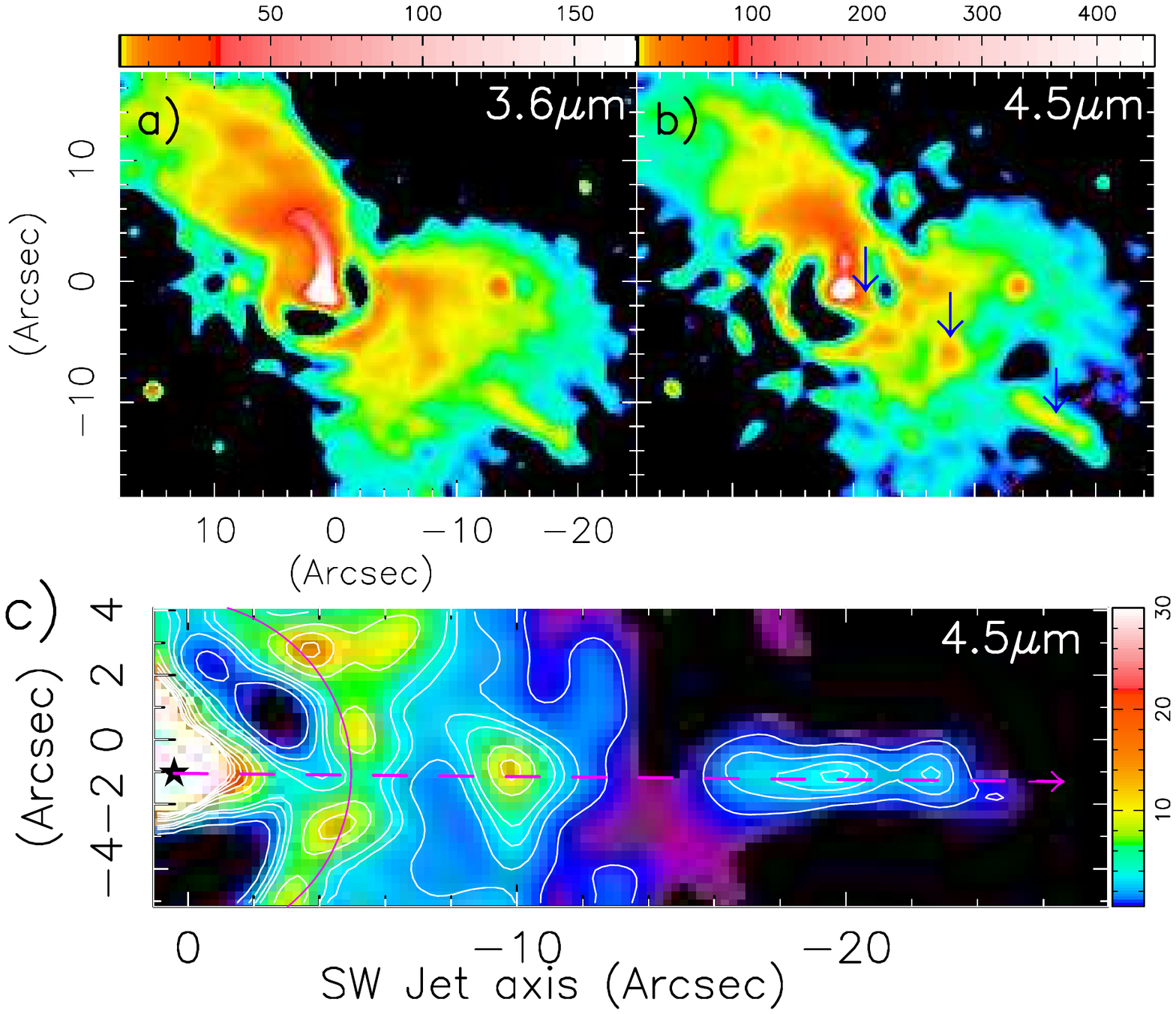} \caption{(a) and (b) HiRes
deconvolved images of the region around the protostar (HH46 IRS) at
3.6 and 4.5 $\mu$m.  Log color stretch is used with highest
intensities at the 20\% and 4\%, level of the protostar peak
brightness at 3.6 and 4.5 $\mu$m, respectively.  The lowest surface
brightness (indicated by blueish green) traces the wide angle
outflows to the NE and SW.  The black arrows in (b) mark the
jet-like protrusion, H$_2$ knot, and the compact jet  which are well
aligned with the SW jet (Fig. 3). The circular arc marks the residue
in the  airy lobe at a level of 0.04\% of the peak ($1.2\times10^{4}
MJy sr^{-1}$). (c) IRAC 4.5 \micron contour map and the gray scale
image of the jet features along the SW lobe.  The contours are at 1,
2, 3, 4, 8, 12, ....
 32 $MJy sr^{-1}$. The star symbol and dshed lines mark   the
protostar and the jet axis.
 \label{fig2}}
\end{figure}

\begin{figure}
\includegraphics  [angle=-90, scale=0.7] {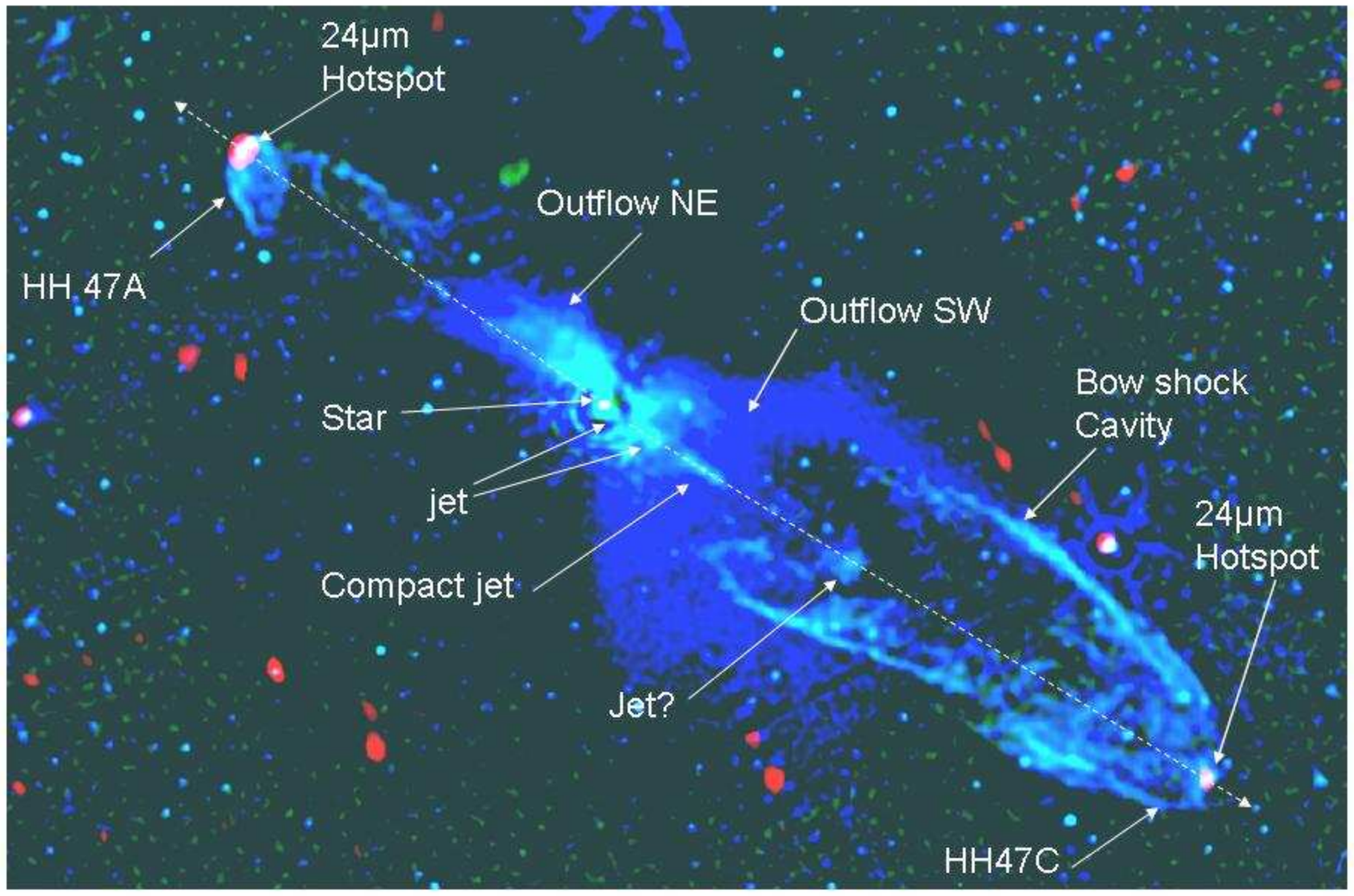}
 \caption{HiRes deconvolved three-color
Spitzer images: IRAC 3.6 \micron (blue), IRAC (4.5\micron +
5.8$\mu$m) (green), MIPS 24 \micron (red).  To avoid confusion, the
diffraction residue around the protostar in the 24 \micron image
inside the circle in Fig.1 is not shown.  The labels identify the
observed features described in the text.
 \label{fig3}}
\end{figure}

\begin{figure}
\epsscale{1.} \plotone{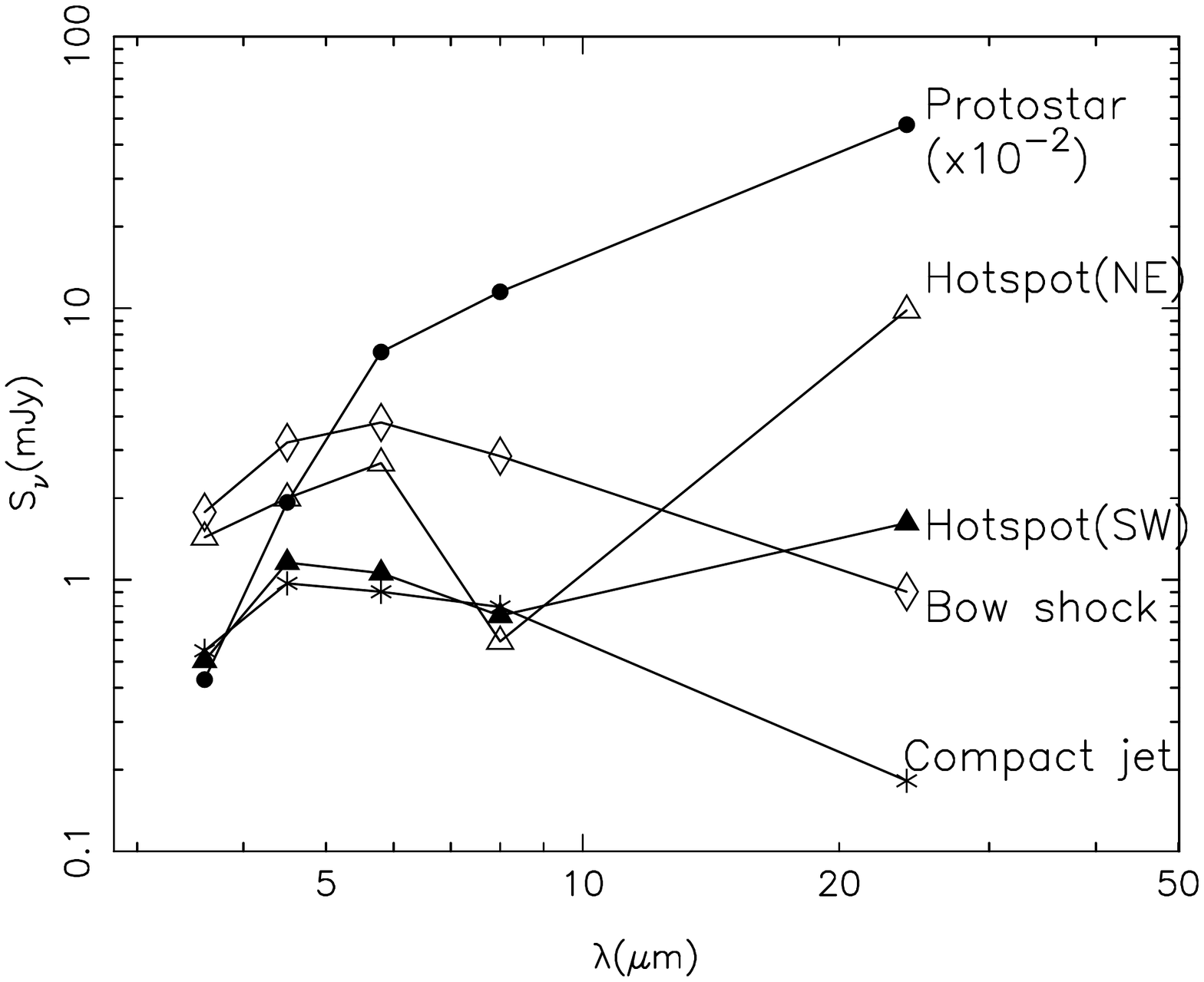} \caption{SEDs at the positions marked
in Fig. 3.  The SNRs  for the fluxes are $>$ 8   at all positions
for all IRAC bands and at the hotspots and protostar for MIPS. The
increase in the SED at 24 \micron for the hotspots is due to [FeII]
lines in this MIPS band.
 \label{fig4}}
\end{figure}
\end{document}